# Thermal conductivity of one-dimensional carbon-boron nitride van der Waals heterostructure: A molecular dynamics study

Han Meng[a], Shigeo Maruyama[b], Rong Xiang[b,] *, Nuo Yang[a,] *

[a] State Key Laboratory of Coal Combustion, School of Energy and Power Engineering, Huazhong University of Science and Technology, Wuhan 430074, China.
[b] Department of Mechanical Engineering, The University of Tokyo, 7-3-1 Hongo, Bunkyo-ku, Tokyo 113-8656, Japan.

* Corresponding author.
E-mail address: xiangrong@photon.t.u-tokyo.ac.jp (R. X.) and nuo@hust.edu.cn (N. Y.)

**ABSTRACT**

Investigating thermal transport in van der Waals heterostructure is of scientific interest and practical importance for their applications in a broad range. In this work, thermal conductivity of one-dimensional heterostructure consisting of carbon and boron nitride nanotubes is systematically investigated via molecular dynamics simulations. Thermal conductivity is found to have strong dependences on temperature, length and diameter. In addition, the axial strain and intensity of van der Waals interaction are demonstrated to be able to modulate thermal conductivity up to about 43% and 37%, respectively. Moreover, the dependence of thermal conductivity on the chirality of componential nanotubes is studied. These results are explained based on lattice dynamics insights. This work not only provides feasible strategies to modulate thermal conductivity, but also enhances the understanding of the fundamental physics of phonon transport in one-dimensional heterostructure.

1. Introduction

The van der Waals (vdW) heterostructure with different atomic layers stacked beyond symmetry and lattice matching through vdW interaction, has been considered as a promising candidate in a wide application range such as electronics, thermoelectrics and photology.[1-6] To the current state of the art, dimensionality of heterostructure is decreased but limited to two-dimensional (2D) form, which is composed of different planar layers like graphene, single-layer boron nitride, phosphorene and single-layer chalcogenides. [7] Most recently, the limitation of dimension has been experimentally broken, generatin g a new class of heterogeneous material named one-dimensional (1D) vdW heterostructure.[8] By combining the advantages of different nanotubes such as carbon nanotube (CNT), boron nitride nanotube (BNNT) and chalcogenide nanotube, it is hopeful to arouse new properties and broaden the application of 1D heterostructures.[9, 10]

Investigating thermal transport in heterostructures is of scientific interest and practical importance for applications like heat dissipation, thermoelectric materials and electronics packaging. [11-13] Tremendous efforts have been devoted to the research on thermal transport in 2D heterostructures over the past years. Liu et al. designed a phosphorene-graphene heterostructure and achieved high interfacial thermal conductance that can be tuned by tensile strain.[11] Ren et al. demonstrated that interfacial thermal resistance is significantly reduced in phononic-mismatched heterostructures and interfacial thermal conductance has a weak temperature dependence, which is different from that of conventional heterostructures achieved by interfacing dissimilar materials.[12] Cai et al. found that BN dominates the thermal transport in BN-silicene bilayer heterostructure due to the weak interaction between two layers.[13] However, study on the thermal transport in 1D heterostructure is lacking so that the thermal properties and physical mechanisms remains to be discovered. Noting that thermal transport can behave abnormally with the decrease of dimension, to be specifically, phonons propagate ballistically in 1D structures rather than diffusively in 2D systems and beyond. [14] Given that, thermal transport in 1D heterostructures is supposed to be different from that in 2D systems. Therefore, there is a great importance and necessity to study thermal conductivity of 1D heterostructures.

In this work, thermal conductivity of 1D heterostructure consisting of CNT and BNNT (denoted as CNT@BNNT) is numerically studied by non-equilibrium molecular dynamics simulation. Firstly, CNT@BNNT with different sizes and different chiral componential nanotubes are constructed. Subsequently, temperature, length and diameter dependence of thermal conductivity are studied. Then, axial strain and intensity of vdW interaction are introduced to modulate thermal conductivity. Lattice dynamics analysis is carried out to explain the strain effect and the influence of the intensity of vdW interaction on thermal conductivity by quantifying phonon density of states. Lastly, the influence of chirality of componential CNT and BNNT on thermal conductivity is studied and the outlook of future research is proposed.

## 2. Model and method

Similar to the double-wall CNT (DWCNT) and double-wall BNNT (DWBNNT), the one-dimensional CNT@BNNT heterostructure (as shown in **Fig. 1c**) is constructed by coaxially nesting a single-wall CNT (SWCNT) into a single-wall BNNT (SWBNNT). The interwall distance of 3.4 Å between inside SWCNT and outside SWBNNT, equaling to the equilibrium interlayer distance of graphene and monolayer boron nitride results from vdW interaction, is generated by the construction procedure. In the following calculation, CNT@BNNT heterostructure consisting of (11, 11) CNT and (16, 16) BNNT is chosen as the smallest structure, where the length and the inside diameter (diameter of inside CNT) are 10 nm and 1.5 nm, respectively. Noting that CNT@BNNT heterostructure is one-dimensional, we only focus on the thermal conductivity along axial direction.

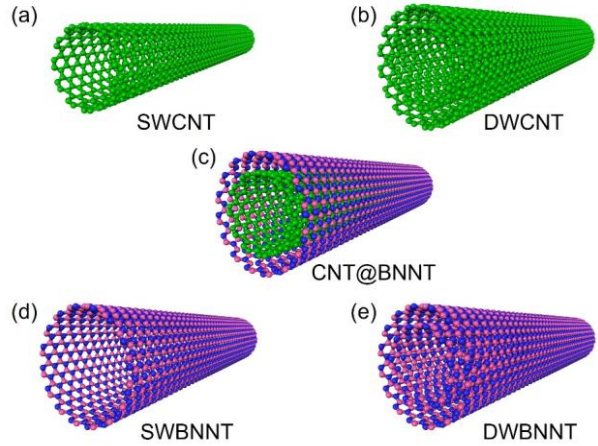

**Fig. 1**. Atomic structure of (a) single-wall CNT, (b) double-wall CNT, (c) 1D vdW heterostructure CNT@BNNT, (d) single-wall BNNT and (e) double-wall BNNT. The pink, blue and green spheres represent boron, nitrogen and carbon atoms respectively.

The classical non-equilibrium molecular dynamics (NEMD) method has been employed in the calculation of thermal conductivity.[15-22] The detailed NEMD simulation setup is illustrated in **Fig. 2a**, where heat source with higher temperature and heat sink with lower temperature are applied to left and right region that are adjacent to the fixed region, respectively. After obtaining the steady temperature profile and heat current (shown in **Fig. 2**), thermal conductivity can be calculated based on the Fourier's law of heat conduction as

$$\kappa = -\frac{J}{A\nabla T}, \tag{1}$$

where $A$ is the cross-sectional area, $\nabla T$ is the time-averaged temperature gradient along axial direction, $J$ is the heat flux that recorded by the average of the input and output power of the two baths as

$$J = \frac{\Delta E_{bath} + \Delta E_{sink}}{2\Delta t}, \tag{2}$$

where $\Delta E$ is the energy flow from heat bath or flow into heat sink during each time step $\Delta t$.

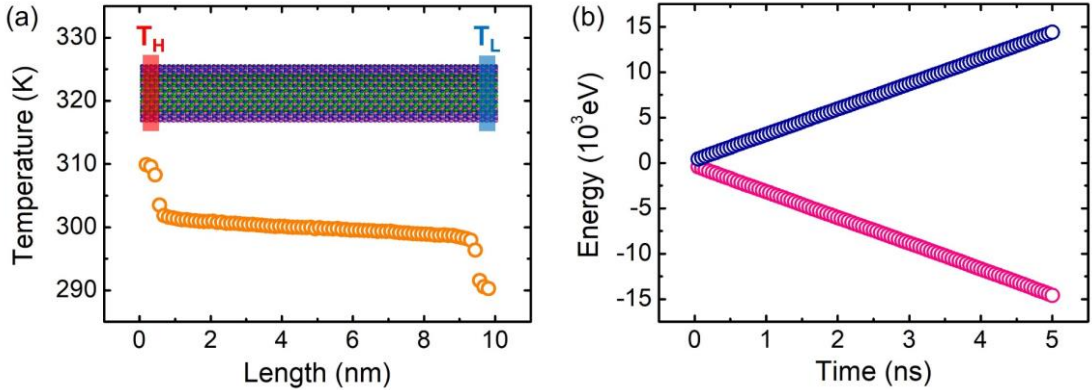

**Fig. 2**. (a) Schematic NEMD simulation setup and time-averaged temperature profile; (b) Accumulated energy change of heat bath and heat sink after the system reaches steady state.

All simulations are performed by the large-scale atomic/molecular massively parallel simulator (LAMMPS) package.[23-28] The interatomic interaction within nanotubes is described by Tersoff potential, which includes both two-body and three-body terms and has been widely used to study the thermal properties.[29, 30] The Lennard-Jones (LJ) potential is implemented for the van der Waals interaction between BNNT and CNT. The LJ parameters are extracted from UFF, parameters across different types of atoms are calculated by using the Lorentz-Berthlot mixing rules.[31] Two Langevin thermostats with a temperature difference of 20 K are used to establish temperature gradient along axial direction. The fixed and periodic boundary conditions are applied in axial and other two directions, respectively. Time step is set as 0.5 fs, and the velocity Verlet algorithm is used to integrate the discrete differential equations of motion.[32] To overcome the statistical error, the results are averaged over five independent simulations with different initial conditions. (More simulation details are given in **Supplementary material**)

3. **Results and discussion**

Firstly, thermal conductivity of CNT@BNNT is calculated at the temperature of 300 K, where the length and the inside diameter are 10 nm and 1.5 nm, respectively. As a result, room-temperature thermal conductivity is obtained as high as 371.61±7.03 $Wm^{-1}K^{-1}$. The value is at the same order of magnitude as other nanotubes such as CNT and BNNT. For a better comparison, thermal conductivity of SWBNNT and SWCNT extracted from CNT@BNNT, DWBNNT and DWCNT and BNNT@CNT with the same size as CNT@BNNT are calculated, the results are shown in **Fig. 3a**. Thermal conductivity of CNT@BNNT is higher than the value of SWBNNT (277.83±5.67 $Wm^{-1}K^{-1}$) but lower than that of SWCNT (553.84±14.14 $Wm^{-1}K^{-1}$). It is reasonable to obtain an intermediate thermal conductivity value when combining two different materials, which reflect the compromise of thermal conductivity of two different materials. Similarly, thermal conductivity of CNT@BNNT also falls in between the values of DWBNNT (271.48 ± 11.28 $Wm^{-1}K^{-1}$) and DWCNT (688.99 ± 21.76 $Wm^{-1}K^{-1}$). Furthermore, thermal conductivity of CNT@BNNT is lower than that of BNNT@CNT (457.84±22.35 $Wm^{-1}K^{-1}$), which is also between the values of corresponding SWBNNT (252.04±18.12 $Wm^{-1}K^{-1}$) and SWCNT (609.98±31.92 $Wm^{-1}K^{-1}$), due to the higher thermal conductivity of larger SWCNT.

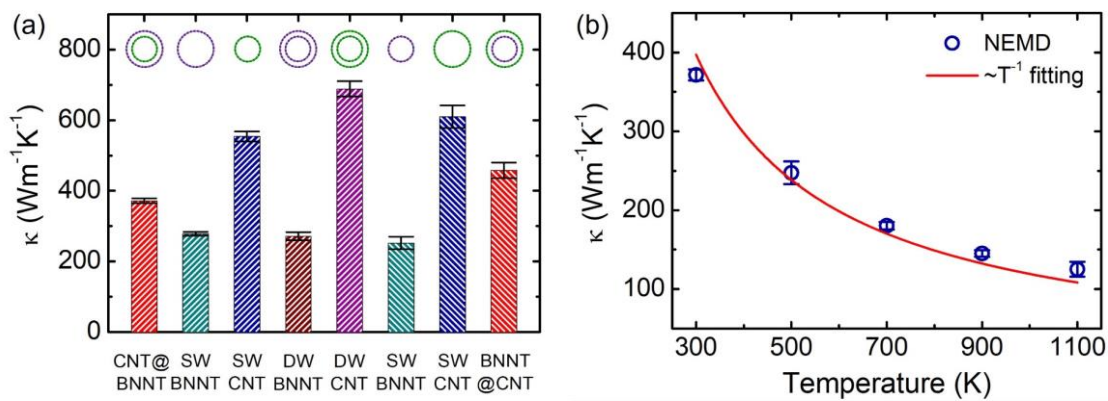

**Fig. 3**. (a) The comparison of thermal conductivity of CNT@BNNT and several nanotubes (SWBNNT, SWCNT, DWBNNT, DWCNT and BNNT@CNT) with the same length of 10 nm at room temperature; (b) The temperature dependence of thermal conductivity of CNT@BNNT. the length and the inside diameter are 10 nm and 1.5 nm, respectively.

Temperature always acts as a factor to significantly affect thermal transport in materials by dominating phonon scattering. Then, temperature dependence of thermal conductivity is further investigated. Thermal conductivity of CNT@BNNT is calculated in the temperature range from 300 to 1100 K, where the length and the inside diameter are fixed at 10 nm and 1.5 nm, respectively. As shown in **Fig. 3b**, thermal conductivity significantly decreases with the increase of temperature, and it drops more than 66% down to $125.12 \pm 9.42$ $Wm^{-1}K^{-1}$ when temperature increases to 1100 K. Moreover, thermal conductivity of CNT@BNNT decreases proportional to the inverse of temperature and can be well fitted, as shown by the red profile. Such a temperature dependence of thermal conductivity can be attributed to the enhancement of Umklapp phonon-phonon scattering that hinder thermal transport at higher temperature. Similar temperature dependence was also found in previous studies on thermal conductivity of other nanotubes and materials.

Size effect of thermal conductivity occurs when system size is reduced to nanoscale.[33, 34] So it is necessary to study size dependence of thermal conductivity for CNT@BNNT. Firstly, length dependence of thermal conductivity is studied. Thermal conductivity of CNT@BNNT is calculated in the length range from 10 nm to 160 nm at 300 K, where the inside diameter is fixed at 1.5 nm. The result in **Fig. 4a** shows that thermal conductivity has a strong dependence on length and increases with the increase of the length. With length increases up to 160 nm, thermal conductivity are obtained as high as $861.67 \pm 9.06$ Wm$^{-1}$K$^{-1}$. The result can be well fitted by an exponential function, which indicates that thermal conductivity increases proportional to $L^\beta$ with a $\beta$ value of $0.31\pm0.02$. Similar length dependence of thermal conductivity was also reported in other nanotubes.[35-37] When length is shorter than phonon mean free path, phonon-phonon interaction can be neglected so that phonons transport ballistically. When length is longer than phonon mean free path, phonon-phonon scattering plays a key role in the process of phonon transport. Previous studies have demonstrated that the divergent thermal conductivity is ascribed to the anomalous heat diffusion induced by super diffusive phonon transport.

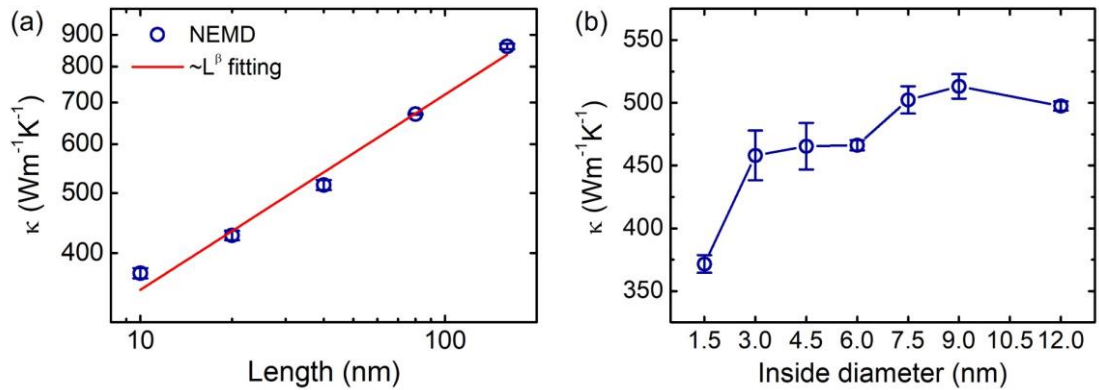

**Fig. 4**. Thermal conductivity of CNT@BNNT as a function of (a) length (the inside diameter is fixed at 1.5 nm) and (b) inside diameter (the length is fixed at 10 nm) at room temperature. The fitting value of $\beta$ is $0.31\pm0.02$.

Secondly, diameter dependence of thermal conductivity is studied. The diameter is increased through magnifying the inside CNT by integral multiple based on the minimum diameter of 1.5 nm, and the distance between inside and outside wall is maintained at 0.34 nm. Thermal conductivity of CNT@BNNT is calculated at different inside diameter at 300 K, where the length is fixed at 10 nm. As shown in **Fig. 4b**, thermal conductivity increases with the increase of inside diameter. Such a diameter dependence of thermal conductivity was also reported in previous study on other nanotubes.[38] There are two aspects mainly account for the increasing thermal conductivity: 1) the number of phonon mode increases with the increase of inside diameter so that more phonons could participate in thermal transport; 2) the increase of inside diameter also leads to the reduction of curvature, which decreases the phonon scattering induced by curve strain. Eventually, thermal conductivity reaches a convergent value around 500 $Wm^{-1}K^{-1}$ when inside diameter is larger than 7.5nm, which is due to the saturation of phonon scattering.

Modulating thermal conductivity is important for thermal management in practical application, as well as understanding the mechanism that governs thermal transport.[39-44] Strain often occurs in a heterostructure as different atomic layers could have distinct coefficients of thermal expansion.[45] This is particularly true in the 1D case as the material is synthesized at high temperatures and, like 2D heterostructures, strain in a nested nanotube cannot be easily released due to it seamless geometry. Naturally, the mechanical strain is introduced to modulate thermal conductivity of CNT@BNNT. Note that the mechanical strain is applied along axial direction of heterostructure. Thermal conductivity of CNT@BNNT with different strain ranges from -0.03 to 0.12 is calculated at 300 K, where the length and the inside diameter are fixed at 10 nm and 1.5 nm, respectively. As shown in **Fig. 5a**, thermal conductivity has a strong dependence on strain, it decreases when nanotube undergoes compression or stretch. When strain decreases to -0.03 under compression, thermal conductivity decreases to 294±4.17 $Wm^{-1}K^{-1}$, which is reduced about 21% compare with the value of primitive nanotube. Thermal conductivity decreases almost linearly with the increase of strain larger than 0.03, and it is reduced about 43% to 210.15±6.88 $Wm^{-1}K^{-1}$ when strain reaches 0.12 under stretch. Previous studies have demonstrated that the strain effect of thermal conductivity is ascribed to the reduction of phonon group velocity, as well as the reduction of phonon relaxation time induced by the increase of the phase space of phonon scattering.[37] This result demonstrates that introducing mechanical strain is an efficient strategy to modulate thermal conductivity of CNT@BNNT.

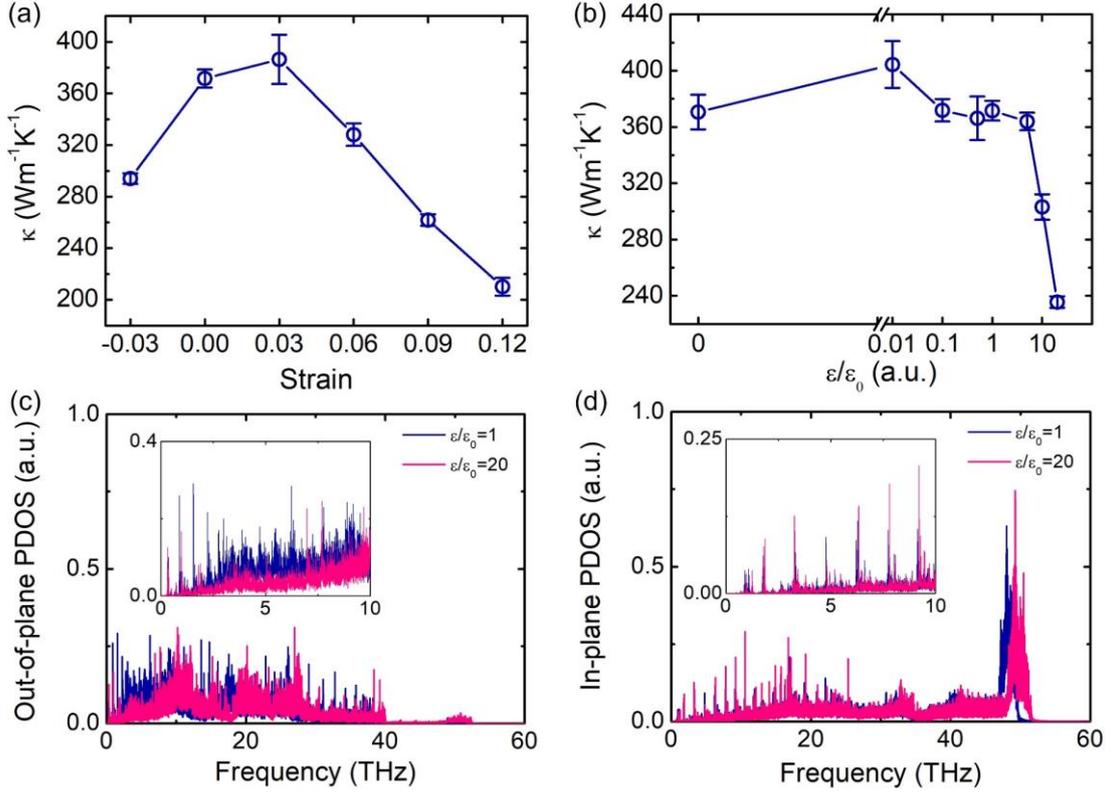

**Fig. 5**. Thermal conductivity of CNT@BNNT as a function of (a) strain along axial direction and (b) normalized intensity of vdW interaction between inside and outside nanotube. The length and the inside diameter are fixed at 10 nm and 1.5 nm respectively; (c) Out-of-plane (radial direction) and (d) in-plane (axial direction) phonon density of states for different intensity of vdW interaction, the insects show the low-frequency region of spectra.

Van der Waals interaction plays an important role between the inside and outside nanotube in heterostructures. Recognizing this, tuning intensity of interaction between two coaxial nanotubes is proposed to modulate thermal conductivity of CNT@BNNT.

The intensity of interaction is adjusted by artificially changing the well depth in Lennard-Jones potential. Here the unitless parameter $\varepsilon/\varepsilon_0$ is defined as the ratio between artificial and original well depth, hence a larger value means more intense interaction and 0 indicates no interaction between inside and outside nanotube. Thermal conductivity of CNT@BNNT is calculated in the ratio range from 0 to 20 at 300 K, where the length and the inside diameter are fixed at 10 nm and 1.5 nm, respectively. As shown in **Fig. 5b**, thermal conductivity has a strong dependence on intensity of interaction. With the ratio increases from 0 to 0.01, thermal conductivity slightly increases, indicating that a weaker inter-layer interaction can slightly facilitate thermal transport. The reason for this facilitation is concluded to that the structural oscillation in out-of-plane that hinder in-plane thermal transport is depressed by the properly slight interaction. Then, thermal conductivity continues to decrease as the ratio increases, especially when the ratio is larger than 5, it has a rapid reduction. When the ratio reaches 20, thermal conductivity decreases 37% and a value of $235.55 \pm 4.07$ $Wm^{-1}K^{-1}$ is obtained. Previous studies have demonstrated that out-of-plane phonons dominate thermal transport in low-dimensional system. To explain the reduction of thermal conductivity, phonon density of states is calculated for both out-of-plane (radial) and in-plane (axial) direction. As shown in **Fig. 5c**, the number of low-frequency range out-of-plane phonons, which contribute more to thermal transport, decreases obviously when the intensity ratio increases from 1 to 20. However, there is no obvious difference in the in-plane PDOS in low-frequency range (shown in **Fig. 5d**). The reduction of thermal conductivity can be attributed to the depression of low-frequency out-of-plane phonons induced by intense interaction between CNT and BNNT. The result indicates that thermal conductivity of CNT@BNNT can be efficiently modulated by tuning intensity of the interaction between two coaxial nanotubes.

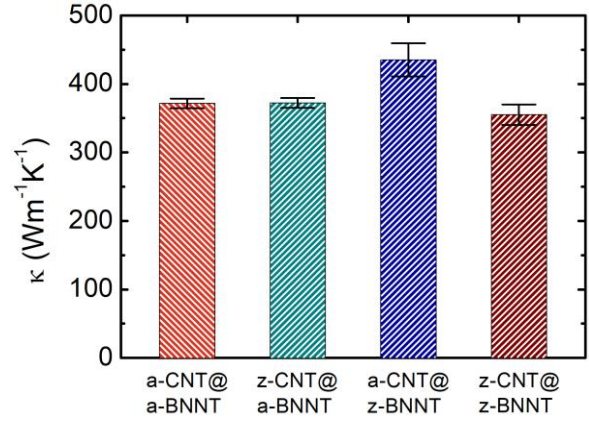

**Fig. 6**. Thermal conductivity of several structures of CNT@BNNT consist of different chiral CNT and BNNT at room temperature. The length and the inside diameter are fixed at 10 nm and 1.5 nm, respectively.

Different chiral single-wall nanotubes are formed through rolling up two-dimensional counterparts along different directions, the structural diversity of CNT@BNNT can be generated through alternating different chiral nanotubes. However, the chirality dependence of thermal conductivity remains blank and needs to be investigated. Due to the great amount of chirality, it is difficult and impossible to investigate all combinations. Therefore, we only focus on armchair and zigzag componential CNT and BNNT (a-CNT, z-CNT and a-BNNT, z-BNNT) without considering other chirality. As shown in **Fig. 6**, thermal conductivity of CNT@BNNT with different chirality combination is calculated at 300 K, where the length and the inside diameter are fixed at 10 nm and 1.5 nm, respectively. The result shows that thermal conductivity of a-CNT@a-BNNT ($371.61 \pm 7.03$ $Wm^{-1}K^{-1}$) and z-CNT@a-BNNT ($372.18 \pm 7.31$ $Wm^{-1}K^{-1}$) are approximately same. Among these four structures, a-CNT@z-BNNT and z-CNT@z-BNNT are calculated to have the highest and lowest thermal conductivity of $435.21 \pm 24.20$ $Wm^{-1}K^{-1}$ and $355.09 \pm 15.02$ $Wm^{-1}K^{-1}$, respectively. The result indicates that variation of the chirality of componential nanotubes can make difference in thermal conductivity. However, the diversity of chirality is considered finitely so that the variation in combination of chiral nanotube is insufficient. More structures with different combination of chiral nanotubes need to be systematically studied in the future, and the highest and lowest thermal conductivity can be discovered by machine learning methodology, such as material informatics.[46, 47] The reason for the structure with highest and lowest thermal conductivity also needs to be revealed as well.

## 4. Conclusion

In this work, thermal conductivity of 1D heterostructure CNT@BNNT is systematically investigated by NEMD simulations. The result shows that thermal conductivity decreases proportional to the inverse of temperature from 300 to 1100 K, which indicates the dominant three-phonon scattering mechanism. Thermal conductivity increases proportional to the power function of length from 10 to 160 nm, and it also has a strong dependence on inside diameter in the range from 1.5 to 12 nm. The axial strain and intensity of interaction between inside and outside nanotubes are demonstrated to be able to modulate thermal conductivity. The dramatical reduction of thermal conductivity by axial strain is attributed to the decrease of phonon group velocity, as well as reduction of phonon relaxation time induced by the increase of the phase space of phonon scattering. The decrease of thermal conductivity induced by tuning intensity of interaction is ascribed to the depression of low-frequency out-of-plane phonons. Moreover, the dependence of thermal conductivity on chirality of componential nanotubes is also studied in the preliminary stage.

All in all, this work provides feasible strategies to modulate thermal conductivity of CNT@BNNT through temperature, size, strain and vdW interaction, which can be used for thermal management. The useful insights into the fundamental mechanisms that govern the thermal conductivity could be generalized to other 1D heterostructures and further facilitate their practical applications in optical and electronic devices.

**Declaration of Competing Interest**

There are no conflicts of interest to declare.

**Acknowledgements**

This work is sponsored by the National Key Research and Development Project of China (2018YFE0127800). The authors thank the National Supercomputing Center in Tianjin (NSCC-TJ) and China Scientific Computing Grid (ScGrid) for providing assistance in computations.

[6]  Y. Cai, G. Zhang, Y.-W. Zhang, Electronic Properties of Phosphorene/Graphene and Phosphorene/Hexagonal Boron Nitride Heterostructures, The Journal of Physical Chemistry C, 119(24) (2015) 13929-13936.

[7]  H. Meng, M. An, T. Luo, N. Yang, 2 - Thermoelectric applications of chalcogenides, in: X. Liu, S. Lee, J.K. Furdyna, T. Luo, Y.-H. Zhang (Eds.) Chalcogenide, Woodhead Publishing, 2020, pp. 31-56.

[8]  R. Xiang, T. Inoue, Y. Zheng, A. Kumamoto, Y. Qian, Y. Sato, M. Liu, D. Tang, D. Gokhale, J. Guo, K. Hisama, S. Yotsumoto, T. Ogamoto, H. Arai, Y. Kobayashi, H. Zhang, B. Hou, A. Anisimov, M. Maruyama, Y. Miyata, S. Okada, S. Chiashi, Y. Li, J. Kong, E.I. Kauppinen, Y. Ikuhara, K. Suenaga, S. Maruyama, One-dimensional van der Waals heterostructures, Science, 367(6477) (2020) 537-542.

[9]  M.G. Burdanova, R.J. Kashtiban, Y. Zheng, R. Xiang, S. Chiashi, J.M. Woolley, M. Staniforth, E. Sakamoto-Rablah, X. Xie, M. Broome, J. Sloan, A. Anisimov, E.I. Kauppinen, S. Maruyama, J. Lloyd-Hughes, Ultrafast Optoelectronic Processes in 1D Radial van der Waals Heterostructures: Carbon, Boron Nitride, and MoS2 Nanotubes with Coexisting Excitons and Highly Mobile Charges, Nano Letters, 20(5) (2020) 3560-3567.

[10] Y. Qian, S. Seo, I. Jeon, H. Lin, S. Okawa, Y. Zheng, A. Shawky, A. Anisimov, E.I. Kauppinen, J. Kong, R. Xiang, Y. Matsuo, S. Maruyama, MoS2-carbon nanotube heterostructure as efficient hole transporters and conductors in perovskite solar cells, Applied Physics Express, 13(7) (2020) 075009.

[11] X. Liu, J. Gao, G. Zhang, Y.-W. Zhang, Design of phosphorene/graphene heterojunctions for high and tunable interfacial thermal conductance, Nanoscale, 10(42) (2018) 19854-19862.

[12] K. Ren, X. Liu, S. Chen, Y. Cheng, W. Tang, G. Zhang, Remarkable Reduction of Interfacial Thermal Resistance in Nanophononic Heterostructures, Advanced Functional Materials, 30(42) (2020) 2004003.

[13] Y. Cai, Q.-X. Pei, G. Zhang, Y.-W. Zhang, Decoupled electron and phonon transports in hexagonal boron nitride-silicene bilayer heterostructure, Journal of15